\documentclass[10pt,twocolumn,letterpaper]{article}

\usepackage{cvpr}              

\usepackage{graphicx}
\usepackage{amsmath}
\usepackage{amssymb}
\usepackage{booktabs}
\usepackage{multirow}
\usepackage{makecell}
\usepackage{mwe}
\usepackage[table]{xcolor}
\usepackage{cuted}
\usepackage{capt-of}

\usepackage[pagebackref,breaklinks,colorlinks]{hyperref}

\usepackage[capitalize]{cleveref}
\crefname{section}{Sec.}{Secs.}
\Crefname{section}{Section}{Sections}
\Crefname{table}{Table}{Tables}
\crefname{table}{Tab.}{Tabs.}

\def\cvprPaperID{1757} 
\def\confName{CVPR}
\def\confYear{2023}

\begin{document}

\title{Consistent Direct Time-of-Flight Video Depth Super-Resolution}

\author{Zhanghao Sun$^1$
\\
$^1$Stanford University, \href{mailto:zhsun@stanford.edu}{zhsun@stanford.edu} \and
Wei Ye$^2$,
Jinhui Xiong$^2$, 
Gyeongmin Choe$^2$, 
Jialiang Wang$^3$, 
Shuochen Su$^2$, 
Rakesh Ranjan$^2$ \\
$^2$ Meta Reality Labs, $^3$Meta Research}

\maketitle

\begin{strip}\centering
\vspace{-15mm}
\includegraphics[width=\textwidth]{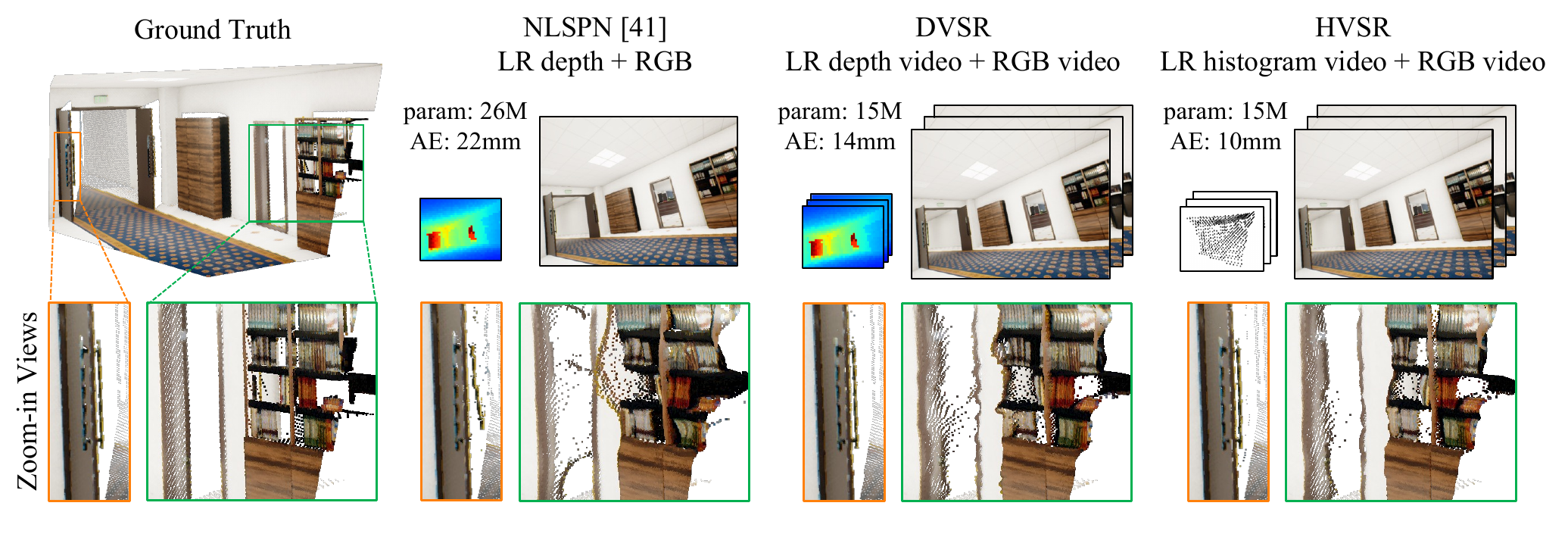}
\vspace{-8mm}
\captionof{figure}{We propose the first multi-frame approaches, dToF depth video super-resolution (DVSR) and histogram video super-resolution (HVSR), to super-resolve low-resolution dToF sensor videos with the high-resolution RGB frame guidance. The point cloud visualizations of depth predictions reveal that, by utilizing multi-frame correlation, DVSR predicts significantly better geometry compared to state-of-the-art per-frame depth enhancement networks~\cite{nlspn} while being more lightweight; HVSR further improves the fidelity of geometry and reduces flying pixels by utilizing the dToF histogram information. Besides the improvements in per-frame estimation, we highly recommend readers to check out the supplementary video, which visualizes the significant improvements in temporal stability across the entire sequences.
\label{fig:teaser}}
\vspace{-3mm}
\end{strip}


\begin{abstract}
\vspace{-5mm}
Direct time-of-flight (dToF) sensors are promising for next-generation on-device 3D sensing. However, limited by manufacturing capabilities in a compact module, the dToF data has a low spatial resolution (\eg $\sim 20\times30$ for iPhone dToF), and it requires a super-resolution step before being passed to downstream tasks. In this paper, we solve this super-resolution problem by fusing the low-resolution dToF data with the corresponding high-resolution RGB guidance. Unlike the conventional RGB-guided depth enhancement approaches, which perform the fusion in a per-frame manner, we propose the first multi-frame fusion scheme to mitigate the spatial ambiguity resulting from the low-resolution dToF imaging. In addition, dToF sensors provide unique depth histogram information for each local patch, and we incorporate this dToF-specific feature in our network design to further alleviate spatial ambiguity. To evaluate our models on complex dynamic indoor environments and to provide a large-scale dToF sensor dataset, we introduce DyDToF, the first synthetic RGB-dToF video dataset that features dynamic objects and a realistic dToF simulator following the physical imaging process. We believe the methods and dataset are beneficial to a broad community as dToF depth sensing is becoming mainstream on mobile devices. Our code and data are publicly available. \url{https://github.com/facebookresearch/DVSR/}
\end{abstract}


\section{Introduction}
\label{sec:intro}

On-device depth estimation is critical in navigation~\cite{slam1}, gaming~\cite{game1}, and augmented/virtual reality~\cite{arkit1, arkit2}. Previously, various solutions based on stereo/structured-light sensors and indirect time-of-flight sensors (iToF)~\cite{kinect, realsense, itoffusion, itof_align} have been proposed. Recently, direct time-of-flight (dToF) sensor brought more interest in both academia~\cite{spad1, spad2, spad3, spad4} and industry~\cite{iphone}, due to its high accuracy, compact form factor, and low power consumption~\cite{dtof_review, dtof3}. However, limited by the manufacturing capability, current dToF sensors have very low spatial resolutions~\cite{dtof_review, dtof_simulator, hndr}. Each dToF pixel captures and pre-processes depth information from a \textit{local patch} in the scene (Sec.~\ref{sec:ifm}), leading to high spatial ambiguity when estimating the high-resolution depth maps for downstream tasks~\cite{arkit1}. Previous RGB-guided depth completion and super-resolution algorithms either assume high resolution spatial information (e.g. high resolution sampling positions)~\cite{s2d1, s2d2} or simplified image formation models (e.g. bilinear downsampling)~\cite{depth_sr1, depth_sr2}. Simple network tweaking and retraining is insufficient in handling the more ill-posed dToF depth super-resolution task. As shown in Fig.~\ref{fig:teaser}, 2nd column, the predictions suffer from geometric distortions and flying pixels. Another fundamental limitation of these previous approaches is they focus on single-frame processing, while in real-world applications, the depth estimation is expected in video (data-stream) format with certain temporal consistency. Processing an RGB-depth video frame-by-frame ignores temporal correlations and leads to significant temporal jittering in the depth estimations~\cite{dont_forget_past, cvd, dcvd}. 


In this paper, we propose to tackle the spatial ambiguity in low-resolution dToF data from two aspects: with information aggregation between multiple frames in an RGB-dToF \textit{video} and with dToF histogram information.
We first design a deep-learning-based RGB-guided dToF video super-resolution (DVSR) framework (Sec.~\ref{sec:depth_video}) that consumes a sequence of high-resolution RGB images and low-resolution dToF depth maps, and predicts a sequence of high-resolution depth maps. Inspired by the recent advances in RGB video processing~\cite{vrt, basicvsrpp}, we loosen the multi-view stereo constraints and utilize flexible, false-tolerant inter-frame alignments to make DVSR agnostic to static or dynamic environments. Compared to per-frame processing baselines, DVSR significantly improves both prediction accuracy and the temporal coherence, as shown in Fig.~\ref{fig:teaser}, 3rd column. Please refer to the supplementary video for temporal visualizations.

Moreover, dToF sensors provide histogram information due to their unique image formation model~\cite{dtof_review}. Instead of a single depth value from other types of 3D sensors, the histogram contains a distribution of depth values within each low-resolution pixel. From this observation, we further propose a histogram processing pipeline based on the physical image formation model and integrate it into the DVSR framework to form a histogram video super-resolution (HVSR) network (Sec.~\ref{sec:hist_video}). In this way, the spatial ambiguity in the depth estimation process is further lifted. As shown in Fig.~\ref{fig:teaser} 4th column, compared to DVSR, the HVSR estimation quality is further improved, especially for fine structures such as the compartments of the cabinet, and it eliminates the flying pixels near edges. 


Another important aspect for deep-learning-based depth estimation models is the training and evaluation datasets. 
Previously, both real-world captured and high quality synthetic dataset have been widely used~\cite{nyuv2, kitti, tartan, hypersim}. However, none of them contain RGB-D video sequences with significant amount of dynamic objects.
To this end, we introduce DyDToF, a synthetic dataset with diverse indoor scenes and animations of dynamic animals (e.g., cats and dogs) (Sec.~\ref{sec:dataset}). We synthesize sequences of RGB images, depth maps, surface normal maps, material albedos, and camera poses. To the best of our knowledge, this is the first dataset that provides dynamic indoor RGB-Depth video. We integrate physics-based dToF sensor simulations in the DyDToF dataset and analyze (1) how the proposed video processing framework generalizes to dynamic scenes and (2) how the low-level data modalities facilitate network training and evaluation. 

In summary, our contributions are in three folds: 

\vspace{-2.0mm}
\begin{itemize}
    \item We introduce RGB-guided dToF video depth super-resolution to resolve inherent spatial ambiguity in such mobile 3D sensor.

    \item We propose neural network based RGB-dToF video super-resolution algorithms to efficiently employ the rich information contained in multi-frame videos and the unique dToF histograms.
    
    \item We introduce the first synthetic dataset with physics-based dToF sensor simulations and diverse dynamic objects. We conduct systematic evaluations on the proposed algorithm and dataset to verify the significant improvements on accuracy and temporal coherence.
\end{itemize}

\section{Related Work}
\label{sec:related}

\subsection{Depth Enhancement Algorithms}
Depth enhancement algorithms convert a degraded depth map into a high-quality one, usually with guidance from a high-quality RGB image~\cite{review_depth_enhance}. They can be roughly divided into two categories: depth completion~\cite{s2d1, s2d2, guidenet, nlspn, penet} and depth super-resolution~\cite{depth_sr1, depth_sr2, depth_sr3}. Depth completion algorithms assume a \textit{high-resolution} depth map with holes or being sparse, and the algorithm inpaints the missing depth by propagating information from the reliable pixels~\cite{nlspn, penet}. Depth super-resolution algorithms accept a \textit{low-resolution} depth map, where each pixel contains mixed information from a patch in the high-resolution ground truth, and the depth information suffers from spatial ambiguity. Lutio et al.~\cite{depth_sr1} model the guided super-resolution process as a pixel-to-pixel mapping and learn it at test-time. The authors further propose a graph-based optimization algorithm to improve the performance~\cite{depth_sr2}. However, they assume the low-resolution depth map is generated with a weighted average sampler (i.e., bilinear downsampling), which is inconsistent with physical image formation models. More importantly, all the previous depth super-resolution research is conducted on a single RGB-D frame. Instead, we aim at dToF video depth super-resolution with the temporal correlation between multiple frames in this paper. We also demonstrate that the utilization of the physical image formation model and the dToF histogram information can further improve performance.


\subsection{Depth Video Processing}
In real-world applications, depth estimations are usually performed on a video (data stream) rather than a single frame. This poses challenges to the temporal stability of the algorithm while also providing at least two additional sources of information: multi-view stereo and temporal correlation between neighboring frames. Previously, lots of efforts have been made to extract the multi-view geometry from a monocular RGB video~\cite{monovideo1, monovideo2, monovideo3, monovideo4} or for self-supervised depth estimation~\cite{dont_forget_past}. However, epipolar constraint does not hold in dynamic environments, and dynamic objects need to be filtered out in the estimation pipeline~\cite{monovideo5}, which limits their applications. On the other hand, how to efficiently utilize the temporal correlation is less explored. 
Patil et al.~\cite{dont_forget_past} use a ConvLSTM structure to fuse concatenated frames without alignment. 
Li et al.~\cite{temp_stereo1} explicitly align multiple frames with a pre-trained scene flow estimator in a stereo video. The performance of these algorithms is largely limited by the inefficient or inaccurate multi-frame alignment module. In this paper, we design a dToF video super-resolution framework with more flexible and false-tolerant multi-frame alignment to better exploit multi-frame correlations.

\subsection{RGB-Depth Dataset}

Synthetic RGB-Depth datasets play an important role in supervised 3D reconstruction tasks due to the high-quality depth ground-truth. They can be generally divided into two categories: indoor environments~\cite{replica, matterport3d, scannet, hypersim, simsin, arkit1} and road environments~\cite{vkitti, apollo, carla}. The current indoor RGB-D dataset generation pipeline usually involves static environmental maps~\cite{hypersim}, and dynamic objects are missing. Road scenes usually contain cars and pedestrians with close to rigid body motions, while their depth distributions are less diverse compared to indoor scenes~\cite{simsin}. To bridge the gap between dynamic real-world indoor environments and static RGB-D datasets, we introduce DyDToF, with animations of animals in static indoor environments.


Apart from the standard RGB color image and depth data modalities, several datasets also simulate specific sensor data based on the physical imaging process. InteriorNet~\cite{interiornet} simulates the event camera and indirect time-of-flight (iToF) sensor outputs. Qi et al.~\cite{flat} simulate the iToF sensor data to handle various common artifacts. In this paper, we simulate the dToF sensor based on the image formation model described in Sec.~\ref{sec:ifm} and provide surface normal maps and material albedos necessary for this simulation.

\begin{figure}[t!]
\begin{center}
\includegraphics[width=0.8\linewidth]{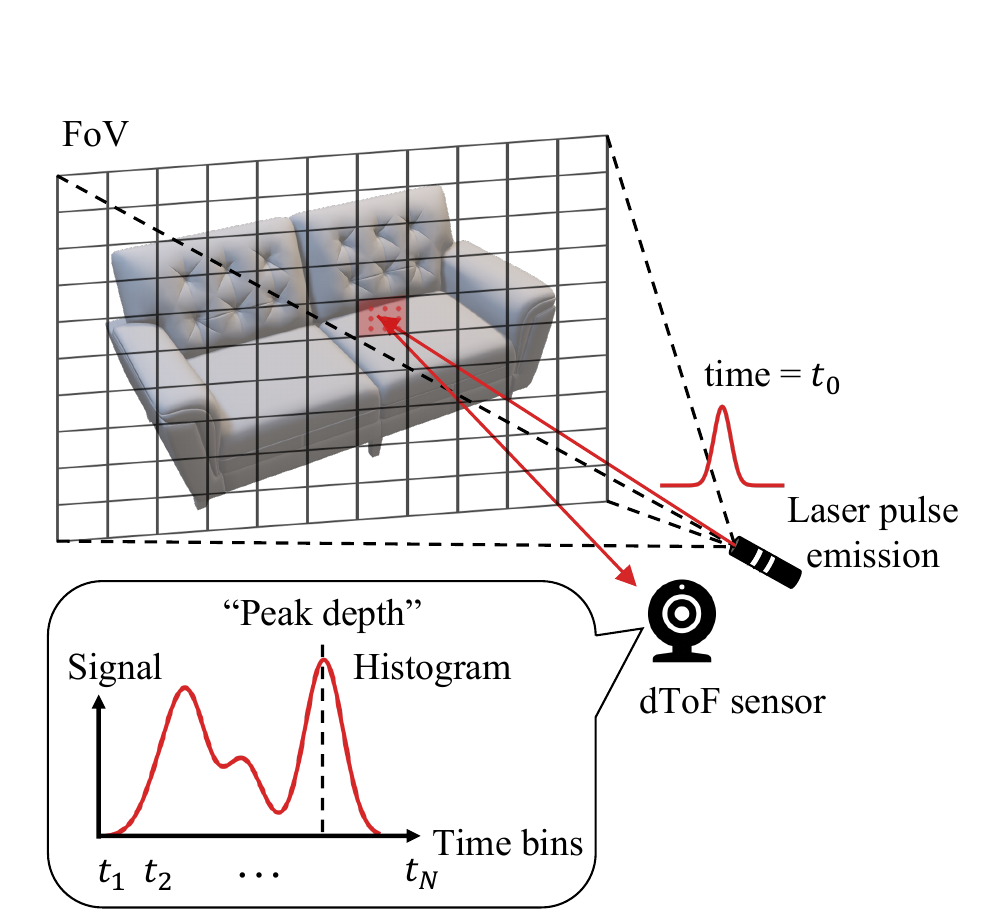}
\vspace{-5mm}
\end{center}
   \caption{Direct time-of-flight (dToF) sensor working principle. Each dToF pixel records a histogram that contains depth information from a patch in the FoV, leading to spatial ambiguity. The dToF sensor can either be operated in ``peak detection'' mode or histogram mode.}
\label{fig:imf}
\end{figure}

\section{dToF Image Formation Model}
\label{sec:ifm}
In this section, we briefly introduce the image formation model for low-resolution dToF sensors, and elaborate on the difference to the previous depth enhancement tasks. Interested readers are referred to~\cite{dtof_review} for more details.

As shown in Fig.~\ref{fig:imf}, a short light pulse is generated by pulsed laser and emitted into the scene. The pulse is scattered and part of the photons will be reflected back to the dToF detector, and triggers arrival events that are time-stamped. According to the time difference between laser emission and receiving, scene depth is determined by the proportional relationship $d = \Delta tc/2$, where $\Delta t$ is the time difference and $c$ is the speed of light. Each dToF pixel captures reflected light from all scene points within its individual field-of-view (iFoV) determined by the overall sensor FoV and the spatial resolution. Therefore, it usually records photon arrival events at multiple time bins. The signal magnitude at each time bin can be expressed as
\begin{eqnarray}
\label{eqn:imf}
\mathbf{h}[k] = \int_{iFoV} \int^{ (k+1)t_0}_{kt_0} r[x,y]g(t - 2d[x,y]/c) dxdydt
\\
\nonumber
k = 1, 2, ..., K
\end{eqnarray}
where $t_0$ is the time bin size, $K$ is the number of time bins (determined by dToF pixel circuitry), $g$ is the laser pulse temporal shape, $d[x,y]$, $r[x,y]$ are the depth and radiance of the scene points within iFoV. We denote the $K$ dimensional signal $\mathbf{h}$ recorded at a single dToF pixel a ``histogram''. We use this image formation model in the following simulations and synthetic data generation (Sec.~\ref{sec:dataset}). 
Similar to conventional depth super-resolution tasks, here we assume the low spatial resolution to be the only degradation in input data. We leave discussions on hardware imperfectness, including shot noise, dark current, depth discretization, and multi-path interference to the supplementary material.

The dToF data can be processed in two modes: ``peak detection'' mode and histogram mode. In the first mode, histogram peak detection is performed at each pixel. Only the peak depth value with strongest signal is sent to the post-processing network. In the second mode, more information contained in the histogram is utilized. In both modes, the dToF data contains relatively accurate depth information, while the lateral spatial information is only known to a low resolution (e.g. $16\times$ lower than desirable). This spatial ambiguity makes the depth super-resolution task significantly more difficult compared to conventional sparse depth completion tasks~\cite{s2d1, s2d2}.

\begin{figure*}
\begin{center}
\includegraphics[width=0.8\linewidth]{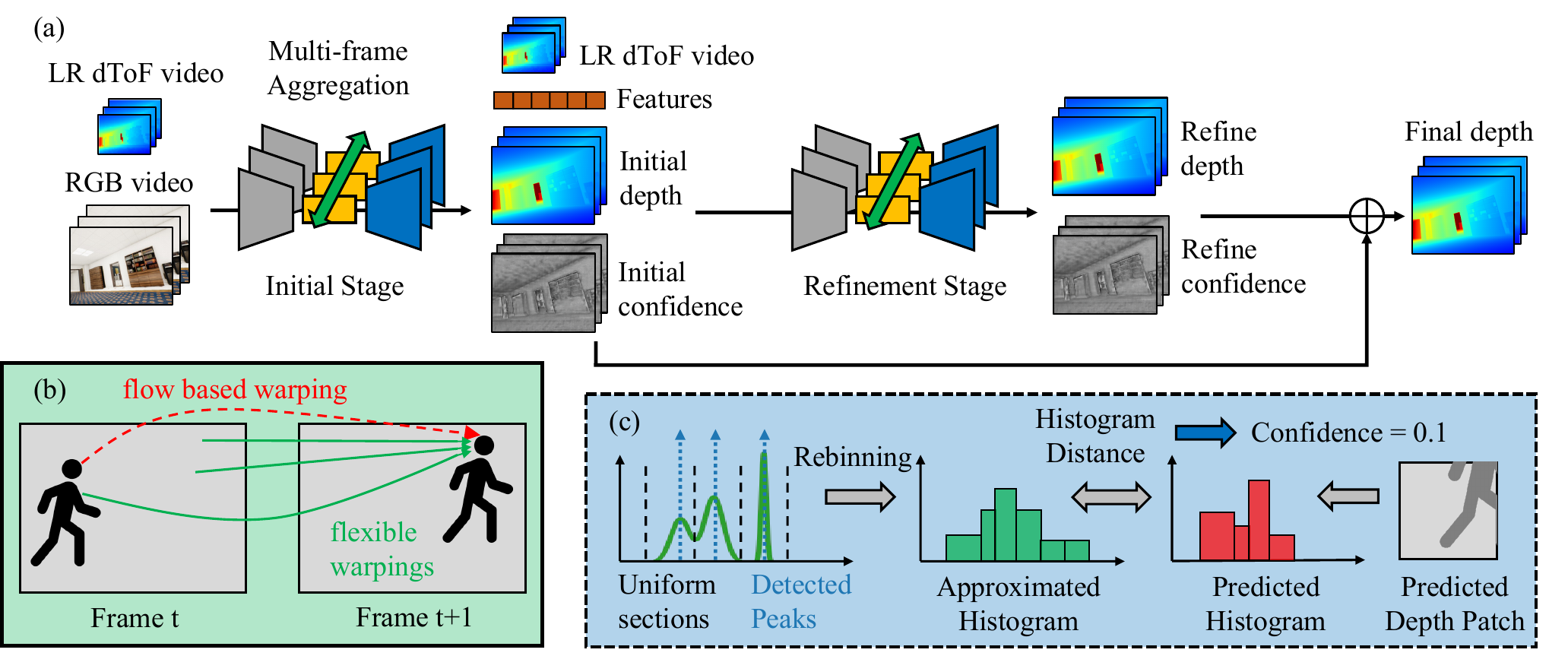}
\vspace{-5mm}
\end{center}
   \caption{(a) Proposed dToF video super-resolution framework. It generally follows a two-stage prediction strategy, where both stages predict a depth map and a confidence map that are fused to obtain the final prediction. Features are aligned and aggregated between frames, either bidirectionally or forward-only. (b) Schematic of flexible warping-based multi-frame feature aggregation. Instead of strictly following the estimated optical flow, features from multiple candidate positions are warped between frames. (c) Schematic of proposed histogram processing pipeline. The full histogram is compressed with peak detection and rebinning to produce an approximated histogram. At the confidence prediction stage, histogram distance is computed between the input histogram and the histogram generated by predicted depth values to estimate confidence in the prediction.}
\label{fig:network}
\vspace{-4mm}
\end{figure*}

\section{Methodology}
The input to our network is a sequence of $T$ frames. Each frame consists of an RGB image with spatial resolution $H\times W$ and a dToF data with spatial resolution $(H/s)\times(W/s)$, where $s$ is the downsampling factor (we use $s=16$ in all experiments). In the histogram mode, the dToF data has an additional temporal dimension with $K$ time bins at each frame, leading to a data volume with dimensions $(H/s)\times(W/s)\times K$. In both modes, our network predicts a sequence of $H\times W$ high-resolution depth maps. 

\subsection{dToF Depth Video Super-Resolution}
\label{sec:depth_video}
The overall RGB-dToF video super-resolution (DVSR) network architecture is shown in Fig.~\ref{fig:network} (a). The network operates in a recurrent manner, where multi-frame information is propagated either forward-only or bidirectionally. 

At each frame, we perform a two-stage processing to predict the high-resolution depth map (same resolution as RGB guidance). In the first stage, the dToF sensor data is fused with the RGB guidance to generate an initial high-resolution depth prediction and a confidence map. The first stage processing results and the dToF sensor data are input into the second stage refinement network to generate the second depth prediction and confidence map. The initial and the second depth predictions are fused according to the confidence maps to generate the final prediction. Apart from the feature extractor and the decoder, each stage contains a multi-frame propagation module and a fusion backbone to sufficiently exchange temporal information and temporally stabilize the depth estimations. Detailed network architecture is provided in the supplementary.

Previous monocular depth video processing algorithms~\cite{monovideo1, cvd, dcvd} usually pose ``hard'' epipolar constraints to extract multi-view geometry. The ``hard'' correspondence search and motion alignment is also employed in processing stereo videos~\cite{temp_stereo1}. Instead, we give the network freedom to pick out multiple helpful correspondences. We jointly finetune a pre-trained optical flow estimator while not posting supervision on the estimated flow. We also include a deformable convolution module after the optical flow-based warping to pick multiple candidates for feature aggregation (as shown in Fig.~\ref{fig:network} (b)). This operation further increases flexibility and compensates for errors in the flow estimations. 
This design choice provides at least two benefits: First, the algorithm can easily generalize to both static and dynamic environments. Second, the correspondence detection between frames does not need to be accurate. Despite the recent progress in deep learning-based approaches, a flow estimator that is both lightweight, fast, and accurate is still missing. Especially, to accurately warp depth values between frames, a 3D scene flow estimation is required, which is more challenging compared to 2D optical flow estimation. State-of-the-art scene flow estimators still suffer from comparatively low accuracy and are limited to rigid body motions~\cite{raft3d}.  

\subsection{dToF Histogram Video Super-Resolution}
\label{sec:hist_video}

Based on the depth video super-resolution network, we further propose a histogram video super-resolution (HVSR) network to employ the unique histogram information provided by dToF sensor. It is impractical to process the full histogram data even with powerful machines. Therefore, simple compression operations are first performed on the temporal dimension of the histogram. Rebinning techniques have been proposed for monocular depth estimations to enforce the network to focus on order relationship~\cite{dorn} and more important depth ranges~\cite{adabins}. As shown in Fig.~\ref{fig:network} (c), here we propose a similar histogram compression strategy: First, we threshold the histogram to remove the signal below the noise floor. Then the histogram is uniformly divided into $M$ sections, and within each section, the peak is detected. We then rebin the histogram into $2M$ time bins defined by the section boundaries and peaks. This $(H/s)\times(W/s)\times M$ data volume is input into the neural network.

We utilize the compressed histogram in two aspects: First, the $M$ detected peaks are concatenated as input to the network in both stages. Second, we compute a histogram matching error to facilitate the confidence predictions. The predicted high resolution depth map $\mathbf{\hat{d}}$ is divided into $s\times s$ patches, each corresponds to one dToF pixel. The depth values within one patch are converted to a histogram $\mathbf{\hat{h}}$ following the image formation model (Eqn.~\ref{eqn:imf}). Then the predicted histogram is compared with the input dToF histogram $\mathbf{h}$. We define the difference between two histograms following the Wasserstein distance~\cite{wdist2}. 
\begin{eqnarray}
\label{eqn:wdist}
\mathcal{D}[\mathbf{h}, \mathbf{\hat{h}}] = ||\mathbf{c} - \mathbf{\hat{c}}||_{1}, \;\;\;
\mathbf{c}[k] = \sum_{i=1}^{k}{\mathbf{h}[k]}
\end{eqnarray}
where $\mathbf{c}$ is the cumulative probability function derived from the histogram $\mathbf{h}$. Larger $\mathcal{D}[\mathbf{h}, \mathbf{\hat{h}}]$ indicates that the predictions within that the corresponding patch is less reliable and should be assigned a lower confidence in refinement. The histogram matching error is input into the confidence prediction layer in both stages of the network. 


\subsection{Implementation Details}
We train the proposed dToF depth and histogram video super-resolution networks on TarTanAir~\cite{tartan}, a large scale RGB-D video dataset. We use 14 out of 18 scenes for training. We simulate the dToF raw data from the ground truth depth map following the image formation model (Eqn.~\ref{eqn:imf}). Since the TarTanAir dataset only provides RGB images, we use the averaged gray scale image to approximate the radiance. We address this issue in the proposed DyDToF dataset for more realistic dToF simulation (Sec.~\ref{sec:dataset}).

We supervise our network with a \textit{per-frame} Charbonnier loss with $\epsilon = 0.01$~\cite{cbloss} and gradient loss. 

\vspace{-6mm}
\begin{eqnarray}
\label{eqn:loss}
\mathcal{L} = \frac{1}{T}\sum_{t=1}^{T}{\sqrt{(\mathbf{d}_t - \hat{\mathbf{d}}_t)^2 + \epsilon^2}} + ||\nabla \mathbf{d}_t - \nabla \mathbf{\hat{d}}_t||_1
\end{eqnarray}

where $\mathbf{d}_t$, $\hat{\mathbf{d}_t}$ are the ground truth and estimated depth maps at frame $t$ and $\nabla$ is the gradient operator.
During training, we divide the long sequences in dataset into shorter ones with $T = 7$ frames. For each video clip, we clip depth values to $[0, 40]$ and normalize them to $[0, 1]$. In all experiments, we set the spatial super-resolution factor $s=16$, and the number of bins in compressed histogram $M=4$. We train our network for a total of roughly $150$k iterations, with batch size $=32$. We use the Adam optimizer~\cite{adam}, with learning rate $1\times10^{-4}$, and a multi-step learning rate decay scheduler with decay rate $0.2$. The training process takes $\sim$ 2 days on $8\times$ Nvidia Tesla-V100 GPUs. 

\begin{table*}
\small
\renewcommand{\arraystretch}{1.2}
  \centering
\begin{tabular}{|c | c || c| c| c | c | c | c|} 
 \hline
 & & \multicolumn{3}{|c|}{TarTanAir Dataset~\cite{tartan}} & \multicolumn{1}{|c|}{Replica Dataset~\cite{replica}} & \multicolumn{1}{|c|}{DyDToF Dataset~\ref{sec:dataset}}
 \\
 \hline
 Methods & Params (M) & AE (mm) $\downarrow$ & $\delta_{1.25} \uparrow$ & TEPE (mm) $\downarrow$ & AE (mm) $\downarrow$ & AE (mm) $\downarrow$ \\ [0.5ex] 
 \hline\hline
 NLSPN~\cite{nlspn} & 26.2 & 48.8 & 0.986 & 26.3 & 30.2 & 35.9 
 \\ 
 \hline
 PENet~\cite{penet} & 131.6 & 58.8 & 0.982 & 29.8 & 27.4 & 29.5
 \\
 \hline\hline
 Per-frame DVSR & 15.5 & 59.2 & 0.981 & 28.5 & 27.6 & 31.2
 \\
 \hline
 DVSR (Ours) & 15.5 & \underline{40.2} & \underline{0.989} & \underline{15.6} & \underline{16.6} & \underline{21.0}
 \\
 \hline
 HVSR (Ours) & 15.5 & \textbf{27.5} & \textbf{0.993} & \textbf{12.8} & \textbf{10.4} & \textbf{12.3}
 \\ [1ex] 
 \hline
\end{tabular}
\vspace{-1mm}
\caption{Quantitative comparisons on TarTanAir, Replica, and DyDToF dataset. Bold font indicates best results and underline indicates second best results. Our network is trained on synthetic TarTanAir dataset with static scenes, while generalize well to real-world captured scenes in Replica dataset and dynamic scenes in DyDToF dataset.}
\label{tab:quantitative}
\vspace{-2mm}
\end{table*}

\section{Results on Public Dataset}
\label{sec:results}

\begin{figure*}[t!]
\begin{center}
\includegraphics[width=0.9\linewidth]{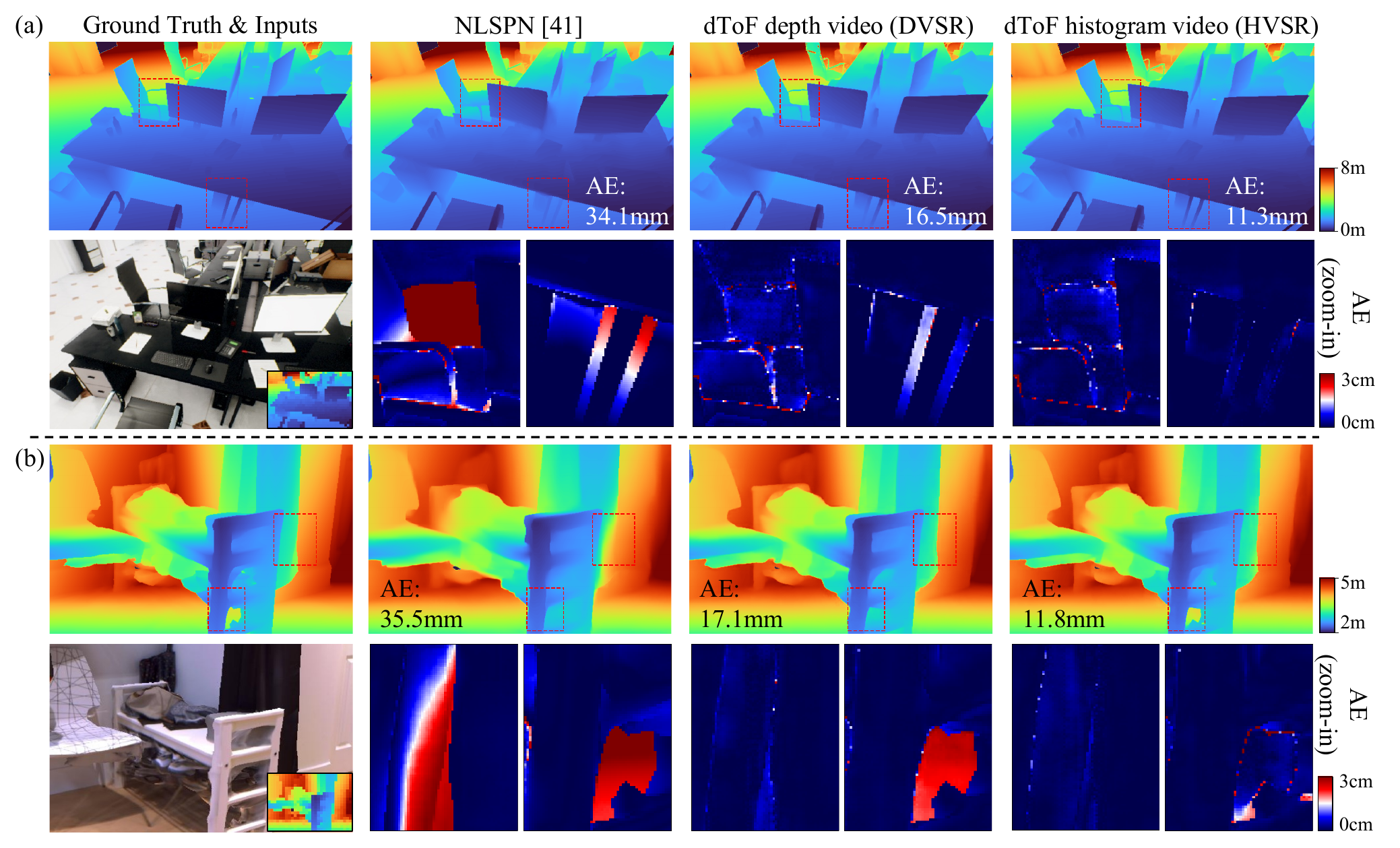}
\vspace{-6mm}
\end{center}
   \caption{Qualitative comparisons on (a) TarTanAir scene, and (b) Replica scene. DVSR and HVSR significantly out-performs the per-frame baseline, especially in zoom-in regions. Please refer to the supplementary video or project page for better temporal visualizations.}
\label{fig:qualitative}
\vspace{-4mm}
\end{figure*}

We evaluate the proposed dToF video super-resolution networks on multiple RGB-D datasets. Since no out-of-the-shelf algorithms directly apply to the dToF sensor super-resolution task, we retrain two state-of-the-art per-frame depth enhancement/completion networks NLSPN~\cite{nlspn} and PENet~\cite{penet}, with the same training settings, as our baselines. As another baseline, we operate the proposed DVSR network in a per-frame manner. We evaluate the depth super-resolution results on three metrics: per-frame absolute error (AE) (lower better), per-frame $\delta_\tau$ metric (higher better), and temporal end-point error (TEPE) (lower better).
\begin{flalign}
\label{eqn:metric}
&\textrm{-- AE (mm): } ||\mathbf{d} - \hat{\mathbf{d}}||_1&
\\
\nonumber
&\textrm{-- } \delta_\tau:\; \textrm{percentage of pixels with max}[d/\hat{d}, \; \hat{d}/d] < \tau &
\\
\nonumber
&\textrm{-- TEPE (mm): } ||(\mathcal{W}[\mathbf{d}_t] - \mathbf{d}_{t+1}) -  (\mathcal{W}[\hat{\mathbf{d}}_t] - \hat{\mathbf{d}}_{t+1})||_1&
\end{flalign}
where $\mathcal{W}$ is the warping operator from frame $t$ to frame $t+1$. We use the ground truth optical flow to perform this warping, and we use the z-buffer aware warping module in PyTorch3D~\cite{pytorch3d} to avoid occlusion induced artifacts. 

\medskip \noindent \textbf{TarTanAir Dataset Evaluation.} We use 4 scenes in TarTanAir dataset, each with $300, 600, 600, 600$ frames, for evaluation. As shown in Table.~\ref{tab:quantitative}, the two video processing networks consistently out-performs the per-frame baselines, despite having less parameters. This validates the effectiveness of multi-frame information aggregation since the proposed network has worse performance when operated per-frame. By utilizing the dToF histogram information, HVSR further boost the estimation quality over DVSR.

We show qualitative comparisons in Fig.~\ref{fig:qualitative} (a). Video processing networks achieve much higher depth qualities compared to the per-frame baselines, especially in fine structures, such as chair arms and thin pillows (better visualized in the zoomed-in bounding boxes). It is evident that fusing information in multiple frames alleviates the spatial ambiguity in processing, since there is high probability that a fine structure not visible in one frame appears in one of its neighboring frames. 

\medskip \noindent \textbf{Replica Dataset Evaluation.} Replica is a real-world captured indoor 3D dataset with realistic scene textures and high quality geometry. We use the same data synthesize pipeline to generate low-resolution dToF data from the ground truth depth and RGB image. We show the cross-dataset generalization of our networks (without fine-tuning) to Replica dataset in Table~\ref{tab:quantitative}, second column. Since ground truth optical flow is unavailable in the Replica dataset, we do not evaluate the temporal metric. We also show qualitative comparisons in Fig.~\ref{fig:qualitative} (b).

\begin{figure}[t!]
\begin{center}
\includegraphics[width=0.9\linewidth]{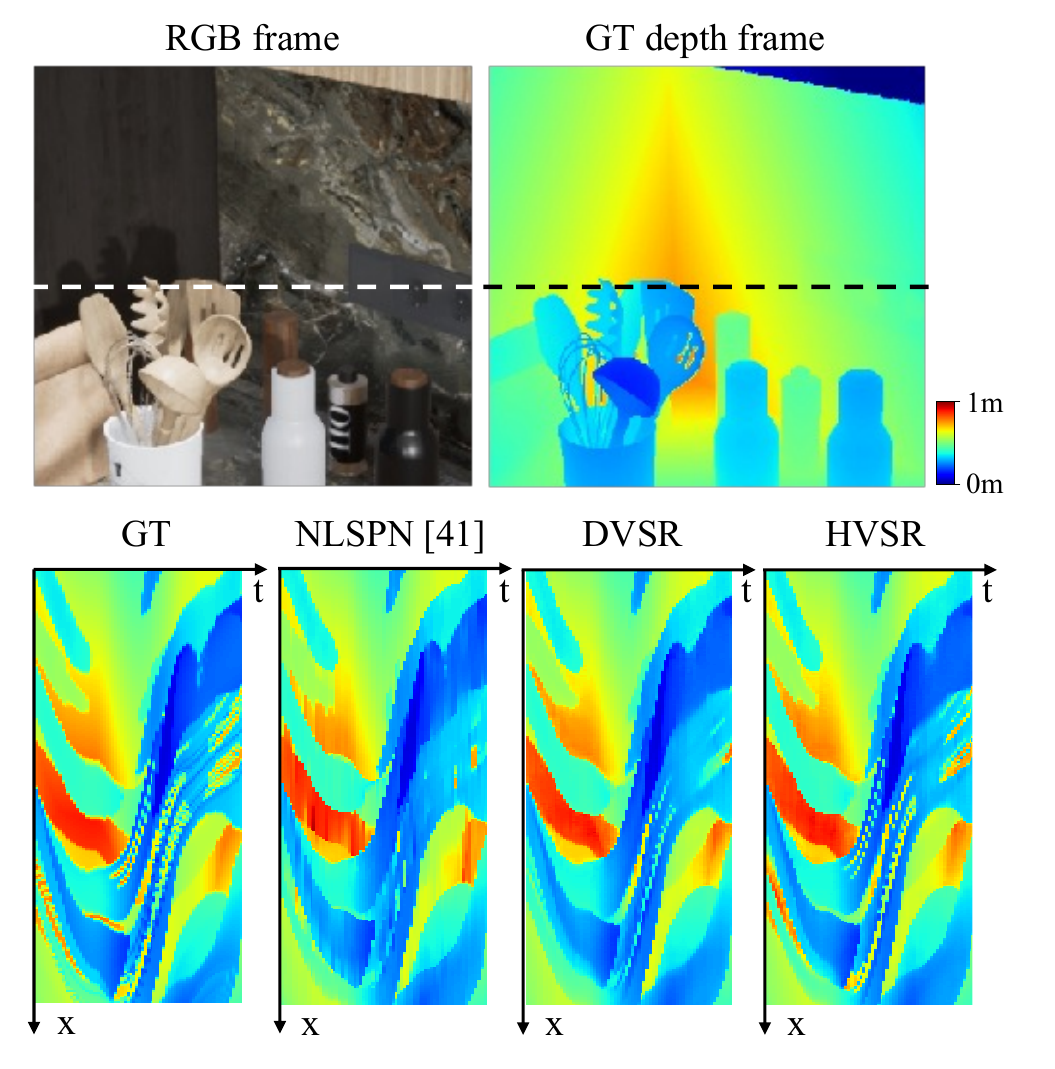}
\vspace{-9mm}
\end{center}
   \caption{x-t slices (along dashed line) for temporal stability visualization. Per-frame baseline has much noiser temporal profile compared to video processing results, while HVSR reveals finer details.}
\label{fig:x-t}
\vspace{-4mm}
\end{figure}

\medskip \noindent \textbf{Temporal Stability.} We also visualize the temporal stability in Fig.~\ref{fig:x-t}, with x-t slices of the estimated depth map. Per-frame processing introduces significant temporal jittering, visualized as noisy/blurred artifacts on the x-t slice. Both DVSR and HVSR has clean x-t slices, demonstrating their high temporal stabilities, while HVSR further reveals fine structures invisible in DVSR predictions. Please refer to the supplementary video or project page for better temporal visualizations.

\begin{figure*}[t!]
\begin{center}
\includegraphics[width=0.9\linewidth]{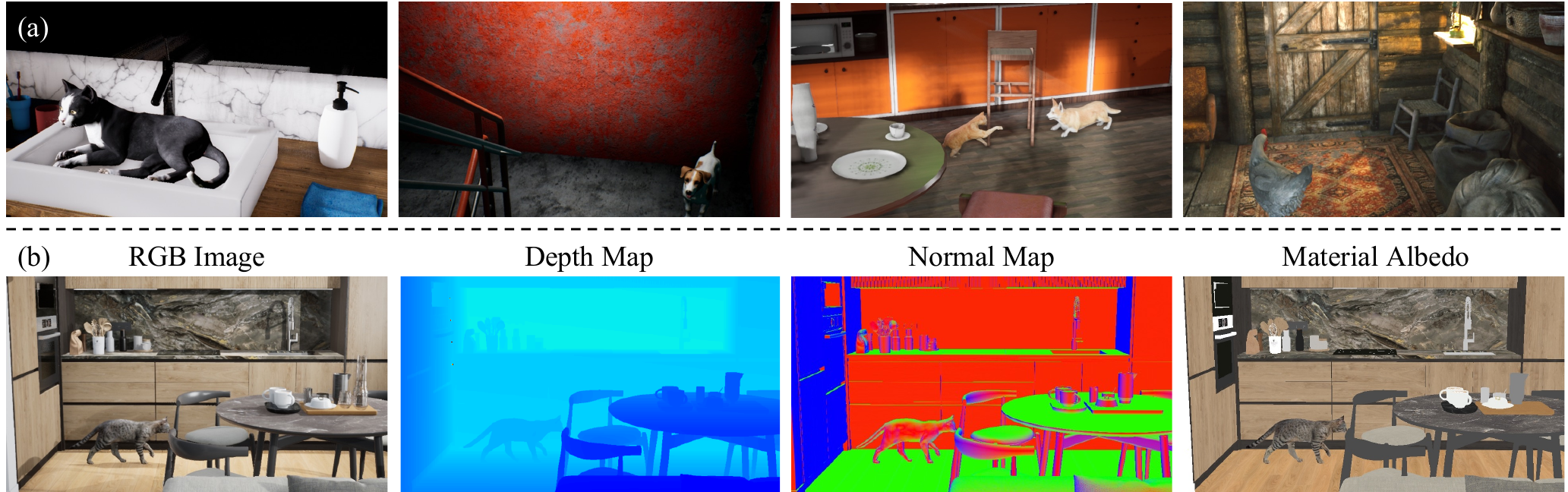}
\vspace{-4mm}
\end{center}
   \caption{DyDToF dataset Overview. (a) We insert dynamic animal models into diverse, high quality indoor environment maps. (b) We generate sequences of RGB images, depth maps, normal maps, material albedos and camera poses.}
\label{fig:dataset}
\vspace{-2mm}
\end{figure*}

\begin{figure*}[t!]
\begin{center}
\includegraphics[width=0.9\linewidth]{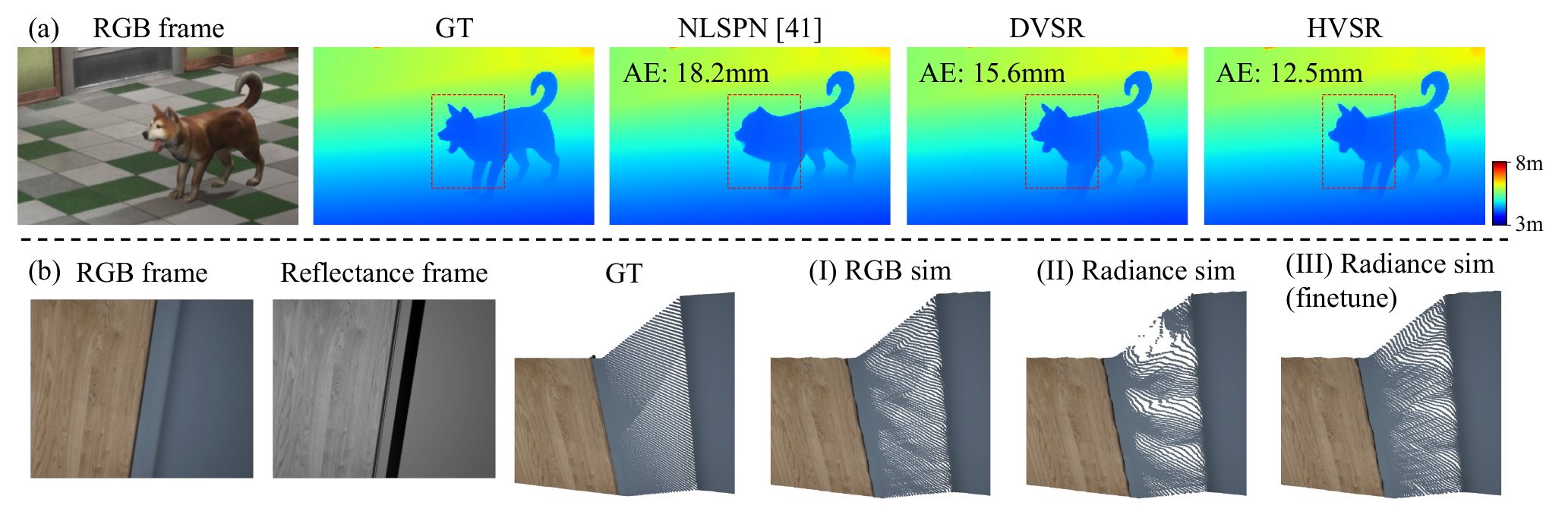}
\vspace{-6mm}
\end{center}
   \caption{Evaluation on DyDToF dataset. (a) The proposed networks DVSR and HVSR performs well with dynamic objects while the per-frame baseline suffers from distortions and blur (b) HVSR trained on TarTanAir dataset fails when there is a mismatch between the RGB image intensity and the radiance computed from physical rendering equation (II). Such artifacts are greatly mitigated by finetuning the network on DyDToF dataset with more realistic dToF simulations (III).}
\label{fig:dataset_eval}
\vspace{-3mm}
\end{figure*}

\section{DyDToF RGB-dToF Video Dataset}
\label{sec:dataset}
Motivated by the lack of dynamic RGB-D video datasets, we introduce DyDToF, with animal animations inserted into indoor environments. An overview of the dataset is shown in Fig.~\ref{fig:dataset}. The dataset contains $100$ sequences ($45$k frames in total) of RGB images, depth maps, normal maps, material albedo, and camera poses generated from Unreal Engine with the open-source plugin EasySynth~\cite{easysynth}. We use $\sim$30 animal meshes (including dogs, cats, birds, and others) associated with $\sim$50 animations in the dataset generation and place them into 20 indoor environments (including schools, offices, apartments, and others). All 3D assets are purchased from publicly available resources.

\subsection{Dynamic Objects Evaluation}
We conduct similar evaluations on the DyDToF dataset and focuses on the depth estimations for dynamic objects. Quantitative comparisons are shown in Table.~\ref{tab:quantitative}, third column. We also show one frame from a barking dog animation in Fig.~\ref{fig:dataset_eval} (a) for qualitative comparison. Although TarTanAir dataset contains very limited amount of dynamic objects, the proposed video networks generalize well to dynamic scenes. We attribute this to our flexible, false-tolerant multi-frame alignment module. Please refer to our supplementary material for ablation studies.

\subsection{More Realistic dToF Simulation}
As mentioned in Sec.~\ref{sec:results}, since TarTanAir dataset does not provide material albedo and surface normals, we approximate the radiance $r$ in the image formation model with RGB image. According to the rendering equation~\cite{rendering}, actual radiance is determined by the material albedo $\alpha$, viewing direction $\mathbf{v}$\footnote{Since we assume the laser and receiver in dToF sensor are co-located, viewing direction is parallel to laser illumination direction.} and surface normal $\mathbf{n}$.

\vspace{-5mm}
\begin{eqnarray}
\label{eqn:imf_albedo}
r = \alpha \; \textrm{max}[\langle\mathbf{n}, \mathbf{v}\rangle, 0] / d^2
\end{eqnarray}

We use this formula in DyDToF dataset to generate more realistic dToF simulations and finetune the networks pretrained on TarTanAir dataset. We show an extreme case in Fig.~\ref{fig:dataset_eval} (b), where one of the side facet of the shelf has very low radiance due to surface normal almost being orthogonal to the dToF laser emission direction. This effect does not exist in the RGB image since the light source is not co-located with the camera. As shown in the 3rd column (I), when the RGB image is used in dToF histogram simulation, pretrained HVSR generalizes well. However, when the physically correct radiance is used in dToF simulation, the pretrained HVSR fails with big geometric distortions (II). By finetuning HVSR on DyDToF, it adapts to a more realistic relationship between captured histogram and underlying geometry and avoids the failure (III). 

\section{Ablation Study on Multi-frame Fusion}
\label{sec:ablation}

We first compare various multi-frame fusion modules, as shown in Table~\ref{tab:mf_fusion}. In the simplest case, features from multiple frames are concatenated without alignment. This significantly reduces the performance since features from unrelated spatial locations are fused together. Flow based alignment use a pretrained (fixed) optical flow estimator to align the features between frames. However, this approach suffers from inaccurate flow estimations and the fundamental problem of foreground-background mixing~\cite{softwarp}. The flexible warping in our proposed framework avoids these issues and give the network freedom to pick out useful information from warped features. Our full multi-frame fusion module utilize bi-directional propagation. However, this forbids the online operation since future information is required. To this end, we replace the bi-directional propagation by a forward-only propagation. As shown in Table~\ref{tab:mf_fusion} third row, this also sacrifices the performance, while it still achieves consistent improvements over the per-frame processing baselines and other inefficient alignment strategies.

\begin{table}
\small
\renewcommand{\arraystretch}{1.2}
  \centering
\begin{tabular}{|c | c| c|} 
 \hline
 DVSR Variants & AE (mm) $\downarrow$ & TEPE (mm) $\downarrow$ \\ [0.5ex] 
 \hline\hline
 w/o alignment & 55.2 & 23.8
 \\
 \hline
 Flow based alignment & 51.4 & 20.9
 \\
 \hline
 Forward only & 44.6 & 19.8
 \\
 \hline
 Full model & 40.2 & 15.6
 \\ [1ex] 
 \hline
\end{tabular}
\caption{Ablation studies on multi-frame fusion module.}
\label{tab:mf_fusion}
\vspace{-5mm}
\end{table}

\section{Discussions \& Conclusions}
\label{sec:conclusion}
\begin{figure}[t!]
\begin{center}
\includegraphics[width=1.0\linewidth]{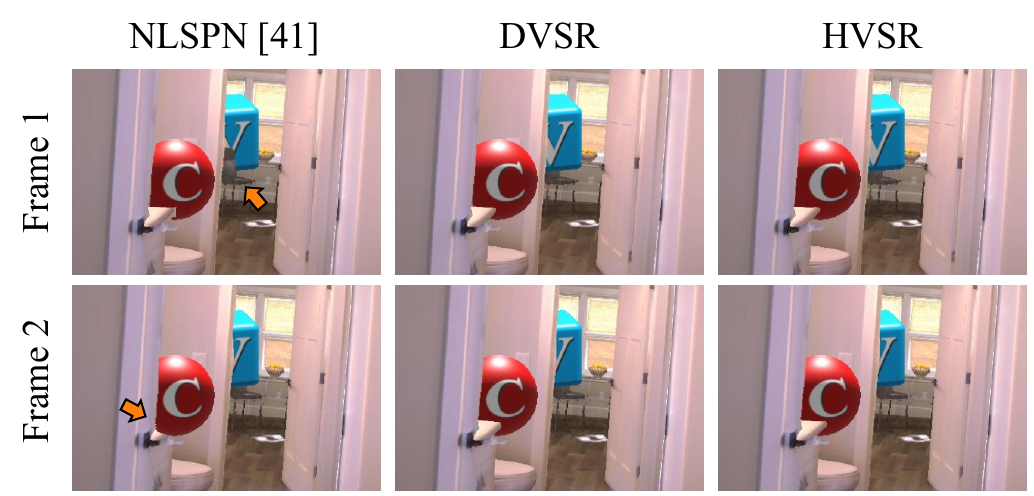}
\vspace{-8mm}
\end{center}
   \caption{Comparison in virtual character insertion (two successive frames). DVSR/HVSR avoids incorrect, temporally inconsistent occlusions in per-frame processing baseline (indicated by orange-colored arrows).}
\label{fig:ar}
\vspace{-6mm}
\end{figure}

In this paper, we propose deep learning frameworks for direct time-of-flight video depth super-resolution. By efficiently employing the multi-frame correlation and histogram information, the proposed algorithms provide more temporally stable and accurate depth estimations for augmented/virtual reality, navigation, and gaming. Especially, the virtual and real environments are fused more seamlessly, leading to better immersive experience. We show one such example, virtual character insertion, in Fig.~\ref{fig:ar} (please refer to supplementary for more results). We also introduce DyDToF, the first indoor RGB-D dataset with dynamic objects, and demonstrate its important role in network training and evaluations. It is not limited to the dToF sensor application and has the potential to set up new benchmarks for general dynamic scene 3D reconstruction and novel view synthesize algorithms~\cite{dcvd, dynamic_3d_1}.

\textit{Envisioning \& Limitations.} (1) The proposed 3D video processing network can be generalized to other types of depth sensors, including stereo and indirect time-of-flight sensors. We show an example of usage in conventional sparse depth completion task in the supplementary material. We leave rest explorations for future research. (2) Limited by the privacy policy of depth sensor vendors, at the time of submission, we do not present evaluations on real-world captured dToF data. However, the DyDToF dataset features commercial-level 3D assets and a physics-based dToF image formation model, thus closes the gap to the real-world application scenario at our best effort.

\section{Acknowledgement}
We would like to thank Yuchen Fan, Xiaoyu Xiang, Greg Cohoon, and Feng Liu for helpful discussions.

{\small
\bibliographystyle{ieee_fullname}
\bibliography{egbib}
}

\end{document}